\documentclass[twocolumn,trackchanges]{aastex701}
\newcommand\lr[1]{{#1}}
\newcommand\lm[1]{{#1}}
\newcommand\lv[1]{{#1}}

\begin{document}

\title{Three-dimensional Magnetic Field Structure of a Quiet-Sun Region Revealed by Sunrise~III/SCIP}

\author[orcid=0000-0001-5616-2808,sname='Kubo']{Masahito~Kubo} \affiliation{National Astronomical Observatory of Japan, 2-21-1 Osawa, Mitaka, Tokyo 181-8588, Japan}\affiliation{Department of Astronomical Science, The Graduate University for Advanced Studies (SOKENDAI), 2-21-1 Osawa, Mitaka, Tokyo 181-8588, Japan}\email{masahito.kubo@nao.ac.jp}

\author[orcid=0000-0002-5054-8782,sname='Katsukawa']{Yukio~Katsukawa} \affiliation{National Astronomical Observatory of Japan, 2-21-1 Osawa, Mitaka, Tokyo 181-8588, Japan}\affiliation{Department of Astronomy, The University of Tokyo, 7-3-1, Hongo, Bunkyo-ku, Tokyo 113-0033, Japan}\affiliation{Department of Astronomical Science, The Graduate University for Advanced Studies (SOKENDAI), 2-21-1 Osawa, Mitaka, Tokyo 181-8588, Japan}\email{yukio.katsukawa@nao.ac.jp}	

\author[orcid=0000-0001-7452-0656,sname='Kawabata']{Yusuke~Kawabata} \affiliation{National Astronomical Observatory of Japan, 2-21-1 Osawa, Mitaka, Tokyo 181-8588, Japan}\email{kawabata.yusuke@nao.ac.jp}	

\author[orcid=0000-0002-7044-6281,sname='Oba']{Takayoshi~Oba} \affiliation{Advanced Research Center for Space Science and Technology, Institute of Science and Engineering, Kanazawa University, Kakuma-machi, Kanazawa, Ishikawa 920-1192, Japan}\affiliation{Max-Planck-Institut für Sonnensystemforschung, Justus-von-Liebig-Weg 3, 37077 Göttingen, Germany}\email{oba@mps.mpg.de}	

\author[orcid=0000-0001-5686-3081,sname='Hara']{Hirohisa~Hara} \affiliation{National Astronomical Observatory of Japan, 2-21-1 Osawa, Mitaka, Tokyo 181-8588, Japan}\email{hirohisa.hara@nao.ac.jp}	

\author[orcid=0000-0003-4764-6856,sname='Shimizu']{Toshifumi~Shimizu} \affiliation{Department of Earth and Planetary Science, The University of Tokyo, 7-3-1, Hongo, Bunkyo-ku, Tokyo 113-0033, Japan}\affiliation{Institute of Space and Astronautical Science, Japan Aerospace Exploration Agency, 3-1-1, Yoshinodai, Chuo-ku, Sagamihara, Kanagawa 252-5210, Japan}\email{shimizu.toshifumi@isas.jaxa.jp}	

\author[orcid=0000-0002-4669-5376,sname='Ishikawa']{Ryohtaroh~T.~Ishikawa} \affiliation{National Institute for Fusion Science, 322-6 Oroshi-cho, Toki City 509-5292, Japan}\email{ishikawa.ryohtaro@nifs.ac.jp}	

\author[orcid=0000-0002-1043-9944,sname='Matsumoto']{Takuma~Matsumoto} \affiliation{Centre for Integrated Data Science, Institute for Space-Earth Environmental Research, Nagoya University, Furocho, Chikusa-ku, Nagoya, Aichi 464-8601, Japan}\email{takuma.matsumoto@gmail.com}	

\author[orcid=0000-0001-6793-8528]{Yoshihiro Naito}
\affiliation{Department of Astronomical Science, School of Physical Sciences, The Graduate University for Advanced Studies, Sokendai, 2-21-1 Osawa, Mitaka, Tokyo, 181-8588, Japan}
\affiliation{National Astronomical Observatory of Japan, 2-21-1 Osawa, Mitaka, Tokyo, 181-8588, Japan}
\email{yoshihiro.naito@grad.nao.ac.jp}

\author[orcid=0009-0005-9709-8431]{Fumihiro Uraguchi}
\affiliation{National Astronomical Observatory of Japan, 2-21-1 Osawa, Mitaka, Tokyo, 181-8588, Japan}
\email{fumihiro.uraguchi@nao.ac.jp}

\author[orcid=0000-0002-8342-8314]{Toshihiro Tsuzuki}
\affiliation{National Astronomical Observatory of Japan, 2-21-1 Osawa, Mitaka, Tokyo, 181-8588, Japan}
\email{toshihiro.tsuzuki@nao.ac.jp}

\author{Kazuya Shinoda}
\affiliation{National Astronomical Observatory of Japan, 2-21-1 Osawa, Mitaka, Tokyo, 181-8588, Japan}
\email{shinoda.kazuya@nao.ac.jp}

\author{Tomonori Tamura}
\affiliation{National Astronomical Observatory of Japan, 2-21-1 Osawa, Mitaka, Tokyo, 181-8588, Japan}
\email{tomonori.tamura@nao.ac.jp}

\author[orcid=0000-0003-4452-858X]{Yoshinori Suematsu}
\affiliation{National Astronomical Observatory of Japan, 2-21-1 Osawa, Mitaka, Tokyo, 181-8588, Japan}
\email{yoshinori.suematsu@nao.ac.jp}

%%% Spain
\author[orcid=0000-0002-3387-026X,sname='del~Toro~Iniesta']{Jose~Carlos~del~Toro~Iniesta} \affiliation{Instituto de Astrofísica de Andalucía, CSIC, Glorieta de la Astronomía s/n, 18008 Granada, Spain}\affiliation{Spanish Space Solar Physics Consortium}\email{jti@iaa.es}	

\author[orcid=0000-0001-8829-1938,sname='Orozco~Suárez']{David~Orozco~Suárez} \affiliation{Instituto de Astrofísica de Andalucía, CSIC, Glorieta de la Astronomía s/n, 18008 Granada, Spain}\affiliation{Spanish Space Solar Physics Consortium}\email{orozco@iaa.es}

\author[orcid=0000-0003-4738-7727,sname='Balaguer~Jiménez']{Maria~Balaguer~Jiménez} \affiliation{Instituto de Astrofísica de Andalucía, CSIC, Glorieta de la Astronomía s/n, 18008 Granada, Spain}\affiliation{Spanish Space Solar Physics Consortium}\email{balaguer@iaa.es}

\author[orcid=0000-0001-5518-8782]{Carlos Quintero Noda}
\affiliation{Instituto de Astrof\'{i}sica de Canarias, E-38205 La Laguna, Tenerife, Spain}
\affiliation{Departamento de Astrof\'{i}sica, Univ. de La Laguna, E-38205 La Laguna, Tenerife, Spain}
\email{carlos.quintero@iac.es}

%%% Principal Investigator:	
\author[orcid=0000-0002-3418-8449,sname='Solanki']{Sami~K.~Solanki} \affiliation{Max-Planck-Institut für Sonnensystemforschung, Justus-von-Liebig-Weg 3, 37077 Göttingen, Germany}\email{solanki@mps.mpg.de}	
	
%%% Lead-CoIs	
\author[orcid=0000-0003-1459-7074,sname='Lagg']{Andreas~Lagg} \affiliation{Max-Planck-Institut für Sonnensystemforschung, Justus-von-Liebig-Weg 3, 37077 Göttingen, Germany}\email{lagg@mps.mpg.de}	
\author[orcid=0000-0002-9972-9840,sname='Gandorfer']{Achim~Gandorfer} \affiliation{Max-Planck-Institut für Sonnensystemforschung, Justus-von-Liebig-Weg 3, 37077 Göttingen, Germany}\email{gandorfer@mps.mpg.de}	

\author[orcid=0000-0002-0787-8954,sname='Bernasconi']{Pietro~Bernasconi} \affiliation{Johns Hopkins University Applied Physics Laboratory, 11100 Johns Hopkins Road, Laurel, Maryland, USA}\email{pietro.bernasconi@jhuapl.edu}	
\author[sname='Berkefeld']{Thomas~Berkefeld} \affiliation{Institut für Sonnenphysik (KIS), Georges-Köhler-Allee 401a, 79110 Freiburg, Germany}\email{thomas.berkefeld@leibniz-kis.de}	
\author[orcid=0009-0009-4425-599X,sname='Feller']{Alex~Feller} \affiliation{Max-Planck-Institut für Sonnensystemforschung, Justus-von-Liebig-Weg 3, 37077 Göttingen, Germany}\email{feller@mps.mpg.de}	
\author[orcid=0000-0001-6317-4380,sname='Riethmüller']{Tino~L.~Riethmüller} \affiliation{Max-Planck-Institut für Sonnensystemforschung, Justus-von-Liebig-Weg 3, 37077 Göttingen, Germany}\email{riethmueller@mps.mpg.de}

%%% Sunrise CoIs	
\author[orcid=0000-0001-9228-3412,sname='Álvarez-Herrero']{Alberto~Álvarez-Herrero} \affiliation{Instituto Nacional de T\'ecnica Aeroespacial (INTA), Ctra. de Ajalvir, km. 4, E-28850 Torrejón de Ardoz, Spain}\affiliation{Spanish Space Solar Physics Consortium}\email{alvareza@inta.es}	
\author[orcid=0000-0003-3490-6532,sname='Smitha']{H.~N.~Smitha} \affiliation{Max-Planck-Institut für Sonnensystemforschung, Justus-von-Liebig-Weg 3, 37077 Göttingen, Germany}\email{narayanamurthy@mps.mpg.de}	
	
\author[sname='Grauf']{Bianca~Grauf} \affiliation{Max-Planck-Institut für Sonnensystemforschung, Justus-von-Liebig-Weg 3, 37077 Göttingen, Germany}\email{grauf@mps.mpg.de}	
\author[sname='Carpenter']{Michael~Carpenter} \affiliation{Johns Hopkins University Applied Physics Laboratory, 11100 Johns Hopkins Road, Laurel, Maryland, USA}\email{michael.carpenter@jhuapl.edu}	
\author[sname='Bell']{Alexander~Bell} \affiliation{Institut für Sonnenphysik (KIS), Georges-Köhler-Allee 401a, 79110 Freiburg, Germany}\email{albe@leibniz-kis.de}	
\author[orcid=0000-0001-7764-6895,sname='Martínez~Pillet']{Valentín~Martínez~Pillet} \affiliation{Instituto de Astrofísica de Canarias, Vía Láctea, s/n, E-38205 La Laguna, Spain}\affiliation{Spanish Space Solar Physics Consortium}\email{vmpillet@iac.es}

%%% Early-Career Scientists	
\author[orcid=0000-0002-7318-3536,sname='Bailén']{Francisco~Javier~Bailén} \affiliation{Instituto de Astrofísica de Andalucía, CSIC, Glorieta de la Astronomía s/n, 18008 Granada, Spain}\affiliation{Spanish Space Solar Physics Consortium}\email{fbailen@iaa.es}	
\author[orcid=0000-0002-2055-441X,sname='Blanco~Rodríguez']{Julian~Blanco~Rodríguez} \affiliation{Universitat de Valencia Catedrático José Beltrán 2, E-46980 Paterna-Valencia, Spain}\affiliation{Spanish Space Solar Physics Consortium}\email{julian.blanco@uv.es}	
\author[orcid=0000-0003-4319-2009,sname='Castellanos~Durán']{Juan~Sebastián~Castellanos~Durán} \affiliation{Max-Planck-Institut für Sonnensystemforschung, Justus-von-Liebig-Weg 3, 37077 Göttingen, Germany}\email{castellanos@mps.mpg.de}	
\author[orcid=0009-0002-6808-5154,sname='Harnes']{Edvarda~Harnes} \affiliation{Max-Planck-Institut für Sonnensystemforschung, Justus-von-Liebig-Weg 3, 37077 Göttingen, Germany}\email{harnes@mps.mpg.de}	
\author[orcid=0000-0001-6029-7529,sname='Hölken']{Johannes~Hölken} \affiliation{Max-Planck-Institut für Sonnensystemforschung, Justus-von-Liebig-Weg 3, 37077 Göttingen, Germany}\email{hoelken@mps.mpg.de}
\author[orcid=0000-0003-1409-1145,sname='Iglesias']{Francisco~A.~Iglesias} \affiliation{Max-Planck-Institut für Sonnensystemforschung, Justus-von-Liebig-Weg 3, 37077 Göttingen, Germany}\affiliation{Grupo de Estudios en Heliofísica de Mendoza, CONICET, Universidad de Mendoza, Boulogne sur Mer 683, 5500 Mendoza, Argentina}\email{iglesias@mps.mpg.de}	
\author[orcid=0000-0003-0175-6232,sname='Siu-Tapia']{Azaymi~L.~Siu-Tapia} \affiliation{Instituto de Astrofísica de Andalucía, CSIC, Glorieta de la Astronomía s/n, 18008 Granada, Spain}\affiliation{Spanish Space Solar Physics Consortium}\email{siu@iaa.es}	
\author[orcid=0000-0003-1483-4535,sname='Strecker']{Hanna~Strecker} \affiliation{Instituto de Astrofísica de Andalucía, CSIC, Glorieta de la Astronomía s/n, 18008 Granada, Spain}\affiliation{Spanish Space Solar Physics Consortium}\email{streckerh@iaa.es}	
\author[orcid=0000-0003-1971-5551,sname='Vukadinović']{Dušan~Vukadinović} \affiliation{Institut für Physik, Universität Graz, Universitätsplatz 5, 8010 Graz, Austria}\affiliation{Max-Planck-Institut für Sonnensystemforschung, Justus-von-Liebig-Weg 3, 37077 Göttingen, Germany}\email{vukadinovic@mps.mpg.de}

%%%
\author[orcid=0000-0003-2038-6062,sname='Ondratschek']{Patrick~A. Ondratschek} \affiliation{Max-Planck-Institut für Sonnensystemforschung, Justus-von-Liebig-Weg 3, 37077 Göttingen, Germany}\email{ondratschek@mps.mpg.de}

%\collaboration{all}{The Terra Mater collaboration}

\begin{abstract}
The balloon-borne stratospheric solar observatory \textsc{Sunrise~iii} successfully completed 6.5 days of observations in July 2024.
One of its focal-plane instruments, the Sunrise Chromospheric Infrared spectroPolarimeter (SCIP), is a slit-scanning spectropolarimeter that simultaneously measures full Stokes profiles of multiple spectral lines in the 850~nm and 770~nm bands. 
SCIP obtained an unprecedented data set of a quiet-sun region near disk center, covering a $58\arcsec \times 58\arcsec$ field of view. 
With an integration time of 10~s per slit position, the scan was completed in 107~minutes without interruption, achieving remarkably stable polarimetric precision of 0.03--0.04\% (1$\sigma$) of the continuum level.
The multi-wavelength SCIP observations reveal that the chromospheric line-of-sight (LOS) magnetic field exhibits thread-like, elongated structures over the internetwork regions, with no obvious photospheric counterpart directly below. 
These threads are typically narrower than $1\arcsec$ and are embedded within the canopy fields extending from the network regions. 
Their LOS field strengths derived from the weak-field approximation are typically 10--20~G weaker than the surrounding canopy. 
In particularly clear cases, the magnetic polarity of the threads is opposite to that of the adjacent canopy. 
These findings suggest that the canopy field is not simply an expanding structure originating from network regions, but instead has a complex three-dimensional configuration containing numerous \lr{localized substructures}. 
These observations provide new constraints on the quiet-sun magnetic topology from the photosphere to \lr{the} chromosphere.
\end{abstract}



\section{Introduction} \label{sec:intro}
The quiet Sun covers the vast majority of the solar surface and harbors a substantial fraction of the unsigned magnetic flux \citep{2004Natur.430..326T}. 
In quiet-sun regions, strong magnetic flux concentrations are located along the network lanes that outline the supergranular boundaries.
%State-of-the-art observations have revealed photospheric magnetic-field properties in internetwork regions surrounded by the network lanes.
The internetwork is pervaded by mixed-polarity magnetic elements on granular or subgranular scales and nearly horizontal fields are detected between these opposite-polarity magnetic elements \citep[e.g.,][]{2008ApJ...672.1237L, 2009A&A...495..607I, 2009ApJ...700.1391M, 2010ApJ...713.1310I, 2010ApJ...723L.149D, 2016A&A...596A...5M}.
This suggests that small-scale flux emergence occurs ubiquitously throughout the quiet Sun. 
Estimates of the emergence rate further indicate that such small-scale events in the internetwork regions can refresh the magnetic flux on the solar surface within roughly a day and can supply the magnetic flux to the surrounding network \citep{2014ApJ...797...49G, 2017ApJS..229...17S}.

Above the photosphere, the chromosphere is dominated by a magnetic ``canopy'' that fans out from the network concentrations and overarches the internetwork \citep{1983SoPh...87...37J, 1990A&A...234..519S}.
The three-dimensional configuration of the canopy fields has been investigated with multi-wavelength \lr{spectropolarimetric} observations in the ultraviolet \citep{2021SciA....7.8406I} and in the visible/near-infrared \citep[e.g.,][]{2020A&A...642A.210M, 2023ApJ...954L..35D, 2024ApJ...965...15K}. 
High-spatial resolution chromospheric images show thread-like fibrils extending from the network into the internetwork, often with highly dynamic evolution. 
The fibrils mostly trace the magnetic field lines of the canopy \citep{2011A&A...527L...8D, 2013ApJ...768..111S,2017ApJS..229...11J}.
The canopies play a fundamental role in coupling magnetic fields between the photosphere, chromosphere, and corona, and may serve as a preferential site for wave propagation, current sheet formation, and magnetic reconnection.
However, most observational studies of chromospheric magnetic structures of canopies or fibrils have targeted active regions (plage or sunspots). 
Comprehensive studies of the weak quiet-sun magnetic fields in the chromosphere, in particular canopy fields above internetwork regions, remain challenging due to limited magnetic sensitivity.

High-precision, multi-wavelength, high-spatial resolution spectropolarimetric data were obtained for a quiet-sun region with the Sunrise Chromospheric Infrared spectroPolarimeter \citep[SCIP:][]{2026arXiv260317929K} onboard the \textsc{Sunrise} balloon-borne observatory \citep{2010ApJ...723L.127S, 2011SoPh..268....1B, 2017ApJS..229....2S} during its third flight \citep[\textsc{Sunrise~iii:}][]{2025SoPh..300...75K, 2026arXiv260607989S}. 
In this Letter, using this unprecedented data set, we investigate the three-dimensional magnetic field structures from the photosphere to the chromosphere.

\section{Observations and Data Reduction} \label{sec:obs}

\begin{figure*}[ht!]
\includegraphics[width=0.95\linewidth]{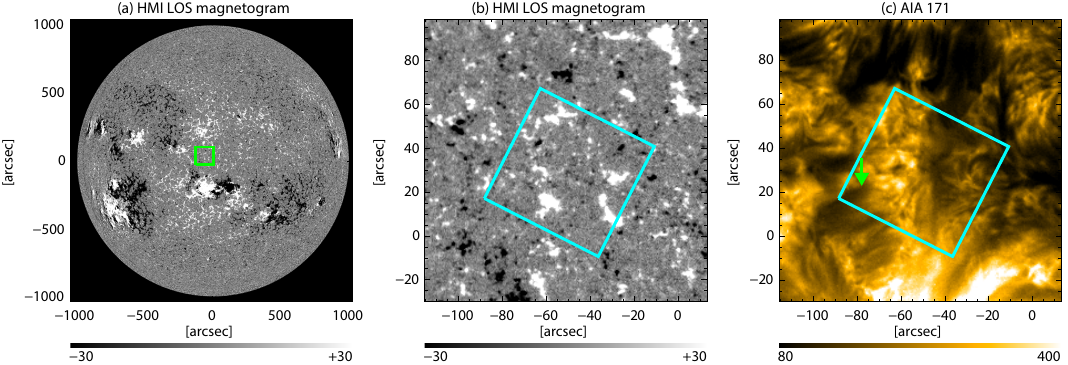}
%\plotone{apjl_SCIP_qs_SDO.eps}
\caption{(\textit{a}) SDO/HMI full-disk LOS magnetogram (Gauss) at 05:38:53~UT on 2024 July~11. 
 The green box indicates the FOV in panels (\textit{b}) and (\textit{c}).
 (\textit{b}) Enlarged SDO/HMI magnetogram of the region in panel (\textit{a}). 
Axes are defined with respect to the disk center along the $X$ and $Y$ directions.
The \lr{cyan} box marks the SCIP FOV.
(\textit{c}) SDO/AIA 171~{\AA} image (DN) at 05:39:33~UT, with the same FOV as panel (\textit{b}). 
Green arrow marks a stable dark coronal feature within the ROI2 FOV (Section~\ref{sec:roi2} for details).
%(\textit{d}) SDO/AIA 304 {\AA} image (DN) at 05:39:29UT. 
\label{fig:SDO}}
\end{figure*}

SCIP performed a single raster scan of a quiet-sun target located near disk center from 04:28:47~UT to 06:15:55~UT on 2024 July~11.
At each slit position, \lm{orthogonal polarization states were recorded simultaneously by two CMOS cameras, one covering a spectral band near 850~nm and the other near 770~nm.}
Five spectropolarimetric exposures of 2.048~s were accumulated onboard and combined into a single data set with a total integration of 10.24~s per position. 
The scan was completed in 107~minutes under very stable conditions, thanks to the excellent performance of the gondola pointing system \citep{2025SoPh..300..112B} and the Correlating Wavefront Sensor \citep{2026SoPh..301...57B}.
The field of view (FOV) is $58\arcsec \times 58\arcsec$ with a pixel scale of $0.094\arcsec$. 
The SCIP FOV in Figure~\ref{fig:SDO} was identified with a full-disk LOS magnetogram from the Helioseismic and Magnetic Imager \citep[HMI;][]{2012SoPh..275..207S} onboard the Solar Dynamics Observatory \citep[SDO;][]{2012SoPh..275....3P}. 
The SCIP slit was rotated by approximately $27^{\circ}$ \lm{clockwise} with respect to solar north. 
Positive magnetic polarity dominates near the disk center, so the target region is not a purely quiet-sun region where magnetic polarities are balanced. 
It may be influenced by a remnant active region in the northern hemisphere and by active region NOAA~13738 in the southern hemisphere near the central meridian. 

We applied the standard calibration pipeline to the SCIP data (Section 5.1 in \citet{2026arXiv260607989S}).
%: dark subtraction, flat-fielding, including intensity correction due to slit-width variation along the slit, correction for spectral dispersion and spectral curvature, wavelength calibration, merging of two orthogonal polarization states, and polarization calibration. 
After calibration, the wavelength sampling is 3.95~pm (3.60~pm) per pixel in the 850~nm (770~nm) band. 
The polarization precision, estimated from continuum wavelengths of Stokes $Q$, $U$, and $V$ data, is 0.04\% ($1\sigma$) in the 850~nm band and 0.03\% ($1\sigma$) in the 770~nm band.
The LOS magnetic field is derived from the SCIP data using the weak-field approximation (Appendix~\ref{sec:wfa}).

\section{Results} \label{sec:res}

\subsection{Overview of the Chromospheric Magnetic Field} \label{sec:overview}

\begin{figure*}[ht!]
\includegraphics[width=0.95\linewidth]{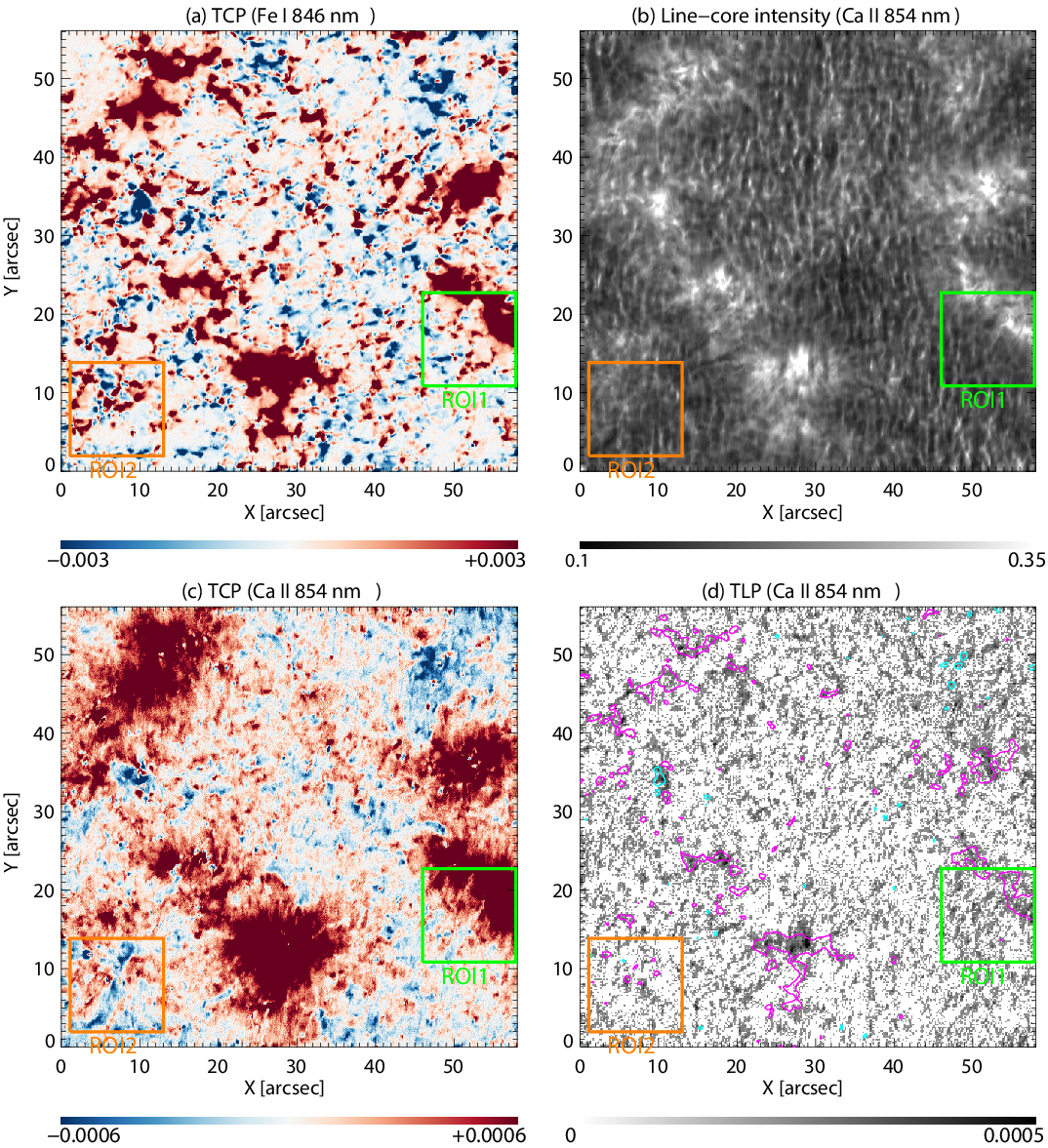}
%\plotone{figures/apjl_SCIP_qs_fFOV.eps}
\caption{
SCIP raster-scan maps of (\textit{a}) the total circular polarization (TCP) of the photospheric \ion{Fe}{1} 846~nm line,  
(\textit{b}) the line-core intensity of the chromospheric \ion{Ca}{2} 854~nm line,
(\textit{c}) TCP of the \ion{Ca}{2} 854~nm line,  
and (\textit{d}) the total linear polarization (TLP) of the \ion{Ca}{2} 854~nm line.
The line-core intensities are normalized to the continuum intensity averaged over the entire FOV.  
The continuum intensity \lv{($I_c$)} is computed by averaging the spectral region from \lv{850.6 to 850.7~nm}.  
Definitions of TCP and TLP are provided in Appendix~\ref{sec:tcp}.  
The TLP map is spatially binned by 2$\times$2 pixels, \lr{and pixels with values below $3\sigma$ of the continuum level are set to zero (white).}
\lr{\lm{The} magenta and cyan contours in panel (d) indicate \lv{TCP (\ion{Fe}{1} 846~nm line) values of $+0.01~I_c$ and $-0.01~I_c$}, \lm{respectively.}}
\lr{The green and orange boxes} indicate the FOVs of Figures~\ref{fig:ROI1} (ROI1) and \ref{fig:ROI2} (ROI2), respectively.
\label{fig:fFOV}}
\end{figure*}

The total circular polarization (TCP, see Appendix~\ref{sec:tcp}) map of the photospheric \ion{Fe}{1} 846~nm line shows small-scale magnetic elements with mixed polarities throughout the internetwork on granular scales (Figure~\ref{fig:fFOV}(a)). 
As discussed in Section~\ref{sec:intro}, this agrees with the previous high-spatial resolution spectropolarimetric studies. 
%A detailed analysis of the photospheric magnetic field is presented in \citet{QuinteroNoda2026InPrep}.
Here we focus on its connection to the chromospheric magnetic field.

The \ion{Ca}{2} 854~nm line-core intensity map shows bright network cores surrounded by dark and bright fibrils that extend into the internetwork (Figure~\ref{fig:fFOV}(b)).
The apparent elongation of structures along the slit direction is particularly prominent in the internetwork. 
This is a projection effect caused by the finite scanning time, effectively tracing the propagation of acoustic waves through the chromosphere. 
The \ion{Ca}{2} 854~nm TCP indicates that positive polarity magnetic fields are predominant both in the photosphere and the chromosphere, while the area occupied by those patches is more extended in the latter atmospheric layers (Figure~\ref{fig:fFOV}(c)). 
This morphology agrees with the expected expansion of the magnetic field with height, thus forming the canopy.
In the internetwork, point-like magnetic elements also exist in the chromosphere, similar to those in the photosphere. 
However, the mixed-polarity fine structure seen in the photosphere is less clear in the chromosphere.

We find that thread-like magnetic field structures are more prominent in the chromosphere.
Narrow features elongated along the vertical (slit) direction likely reflect acoustic-wave patterns, which appear as vertical stripes in the \ion{Ca}{2} 854~nm line-core intensity map. 
In contrast, many threads extend across the slit with widths $\lesssim 1\arcsec$. 
Their persistence across many consecutive slit positions indicates that they are more stable than the vertical stripe patterns.
Near the network boundaries, the threads generally exhibit the same polarity as the network, although some regions (e.g., ROI1 indicated by the green box) contain threads of opposite polarity extending from the network. 
Within the internetwork, numerous negative-polarity threads are also found, including exceptionally long structures highlighted by the \lr{orange} box (ROI2). 
These results suggest that the canopy is highly structured and far from a uniformly expanding fan.

Figure~\ref{fig:fFOV}(d) shows the presence of the linear polarization in the \ion{Ca}{2} 854~nm line in the quiet-sun region.
The strong TLP signals appear \lr{in and around the network, while signals above the noise threshold are also seen widely within the internetwork.}
This morphology likely reflects that, owing to the lower sensitivity to linear polarization compared to circular polarization, only part of the associated magnetic-field structure is detected.
\lv{The linear polarization signals detected here may arise from the Zeeman effect, although the possible contribution of scattering polarization and the detailed interpretation of these signals remain important topics for future investigation.}

\subsection{ROI1: Spine and Intraspine Canopy Magnetic Field}\label{sec:roi1}

\begin{figure*}[ht!]
\includegraphics[width=0.95\linewidth]{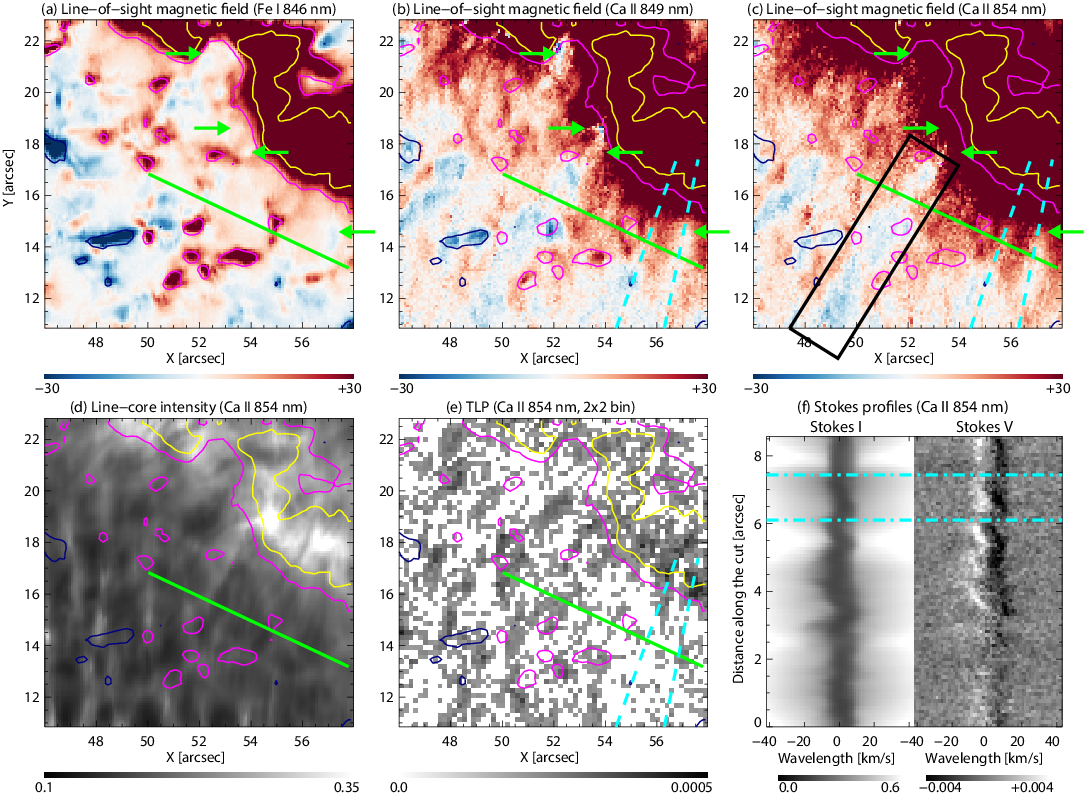}
%\plotone{figures/apjl_SCIP_qs_ROI1_v21.eps}
\caption{
ROI1: 
(\textit{a})--(\textit{c}) LOS magnetic field from the weak-field approximation for \ion{Fe}{1} 846~nm, \ion{Ca}{2} 849~nm, and \ion{Ca}{2} 854~nm.  
(\textit{d}) \ion{Ca}{2} 854~nm line-core intensity.  
(\textit{e}) TLP of the \ion{Ca}{2} 854~nm line (2$\times$2 binning).  
(\textit{f}) Wavelength–space maps along the \lr{green solid} line \lr{(zero at the left end of the green solid line)} for Stokes $I$ (left) and Stokes $V$ (right).  
\lr{\lm{The} magenta, navy, and yellow contours indicate values of $+20$, $-20$, and $+100$~G, respectively.}
\lr{In the TLP map, pixels with values below $3\sigma$ of the continuum level are set to zero (white).}
\lr{\lm{The} cyan} dashed lines trace the weaker LOS magnetic field component of the canopy.
\lr{\lm{The} cyan} dash-dotted lines in panel (\textit{f}) indicate the intersections between the \lr{green and cyan} lines.
\lr{\lm{The} green arrows in panel (a) indicate \lm{weaker photospheric magnetic elements, some of opposite polarity,} located just outside the network region.}
\lr{\lm{The} black box in panel (c) highlights the particularly prominent opposite-polarity thread.}
\label{fig:ROI1}}
\end{figure*}

\begin{figure*}[ht!]
\includegraphics[width=0.95\linewidth]{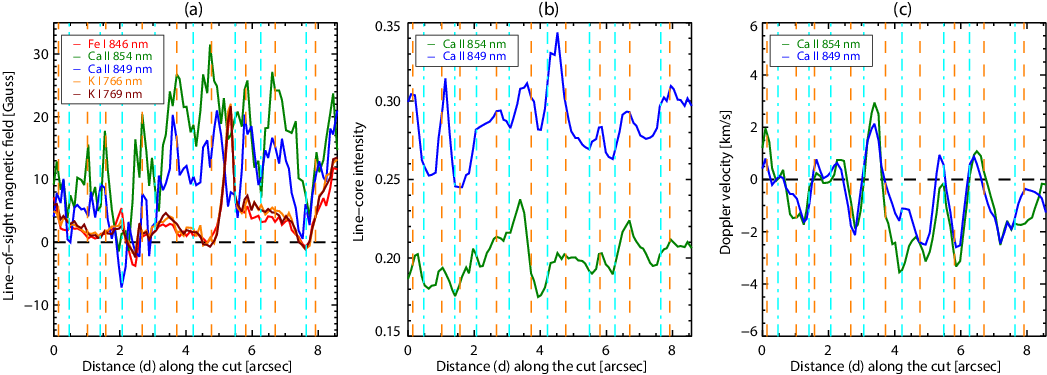}
%\plotone{figures/apjl_SCIP_qs_ROI1_plot.eps}
\caption{
(\textit{a}) LOS magnetic field strength, (\textit{b}) line-core intensity, and (\textit{c}) Doppler velocity along the \lr{green solid} line in Figure~\ref{fig:ROI1}.
\lr{The horizontal axis represents the distance $d$ along the cut, with zero corresponding to the left end of the green solid line.}
\lr{The typical $1\sigma$ uncertainty in the LOS field strength is $\approx4.4$~G (Appendix~\ref{sec:wfa}).}
Definition of the Doppler velocity is provided in Appendix~\ref{sec:dp}.  The positive velocity corresponds to the redshift. The \lr{orange} dashed line and \lr{cyan} dash-dotted line indicate the peak and valley positions of the chromospheric LOS magnetic field (manually identified), respectively.
\label{fig:ROI1_plot}}
\end{figure*}

A detailed examination of the network boundary (ROI1) reveals an alternating pattern of filamentary stronger and weaker LOS magnetic field components along the canopy azimuth in the chromosphere (Figure~\ref{fig:ROI1}(b)--(c)).
This pattern is reminiscent of the spine/intraspine structure of sunspot penumbrae in the photosphere \lm{\citep[][and references therein]{1993A&A...275..283S, 1993ApJ...418..928L, 2011LRSP....8....4B}}.
One particularly prominent feature at the center of ROI1 has opposite polarity to the surrounding network and extends more than 5$\arcsec$ away from the network boundary into the internetwork \lr{(black box in Figure~\ref{fig:ROI1}(c))}.
Although the canopy fields in the \ion{Ca}{2} 854~nm line are slightly more diffuse and extend more into the internetwork compared with those in the \ion{Ca}{2} 849~nm line due to the higher formation height, the alternating pattern is observed in both lines.
\lr{Figure~\ref{fig:ROI1_plot} shows the LOS magnetic field strength, line-core intensity, and Doppler velocity along the green solid line in Figure~\ref{fig:ROI1}. 
The chromospheric LOS field strength fluctuates with amplitudes of about $10$--$20$~G on sub-arcsecond spatial scales.
These fluctuations are 2.3--4.5 times larger than the $1\sigma$ uncertainty of $\approx 4.4$~G estimated from a noise simulation (Appendix~\ref{sec:wfa}).}
The polarity usually remains the same across these fluctuations, but opposite polarities intermittently appear (e.g., at \lr{$d = 2\arcsec$}). 
This opposite-polarity filamentary structure tends to show a slightly larger amplitude of the LOS field fluctuation and a broader spatial width than those in the same polarity.
In the canopy, the LOS field strength and its fluctuation in the \ion{Ca}{2} 854~nm line are slightly larger than those in the \ion{Ca}{2} 849~nm line.
In the photosphere, the LOS magnetic fields show little fluctuation in the same locations, except for spiky changes corresponding to magnetic patches.
These results indicate that the spine and intraspine structures form in the chromosphere and become more prominent with height from the photosphere to the mid-chromosphere.

Enhanced linear polarization signals in the \ion{Ca}{2} 854~nm line, exceeding 5$\sigma$ of the noise level in the continuum, are detected \lr{at the outer boundary of the (photospheric) network} (Figure~\ref{fig:ROI1}(e)).
These signals are located \lr{near the network-side ends} of the weaker components of the LOS magnetic field in the chromosphere as indicated by the \lr{cyan} dashed lines, and they correspond to the intraspine canopy structures.
\lr{In the photosphere, the area immediately outside the network is dominated by \lm{weak magnetic patches with the same polarity}. 
Within this region, \lm{weaker photospheric magnetic elements, some of opposite polarity,} are seen near the network-side ends of the 
intraspine canopy structures (green arrows in Figure~\ref{fig:ROI1}(a)), although the individual photospheric elements are weak.}
%In addition, the weak, small opposite-polarity magnetic elements are located in the photosphere just outside of the network region at the inner ends of the intraspine canopy structures (green arrows in Figure~\ref{fig:ROI1}(a)).

Figure~\ref{fig:ROI1}(f) shows that the Stokes~$V$ amplitude of the \ion{Ca}{2} 854~nm line varies spatially, \lv{which results in spatial variations of the inferred LOS magnetic field.}
In the Stokes~$I$ profile, fluctuations in intensity and Doppler velocity can be seen with similar spatial scales to the Stokes~$V$ amplitude fluctuations. 
Figure~\ref{fig:ROI1_plot} shows \lr{that} the stronger (\lr{orange}-dashed line) and weaker (\lr{cyan} dash-dotted line) LOS magnetic fields tend to \lr{be associated with} relatively brighter and darker chromospheric structures, respectively, although this relationship is not exact at every location.
\lr{For instance, this tendency is better seen at $d < 2\arcsec$ and $d > 5\arcsec$, whereas the peak positions are less well aligned at $2\arcsec < d < 4\arcsec$.}
\lv{This suggests that the LOS magnetic field strength and the line-core intensity are not strongly correlated overall.}
These intensity features appear to stream outward from the network into the internetwork, consistent with the geometry of the canopy (Figure~\ref{fig:ROI1}(d)).
\lr{The Doppler velocity pattern in Figure~\ref{fig:ROI1_plot}(c) \lv{also exhibits local variations on spatial scales comparable to those of the LOS magnetic-field fluctuations.
In some locations, this pattern} is spatially phase-shifted relative to the LOS magnetic field fluctuations, with the velocity peaks tending to lie between the peaks and valleys of the LOS magnetic fields. 
In this sense, each stronger or weaker LOS magnetic field component \lv{can be} flanked by blueshifts on one side and redshifts on the other.
\lv{Such local velocity arrangements} \lm{may suggest} rotational motions \lm{tangential to the axis of} the thread-like structures.
\lv{However, the pattern is local rather than coherent over the full cut, and no strong correlation is found between the Doppler velocity and the LOS magnetic-field strength.}
These tendencies are also seen in the \ion{Ca}{2} 849~nm line.}
%On the other hand, these magnetic field components are not located at the peak positions of the Doppler velocity in the chromosphere, but phase-shifted by on average about 0.5\arcsec. The boundaries between the strong and weak-field components tend to exhibit larger Doppler velocities, with opposite signs on either side.
%On the other hand, the boundaries between the strong- and weak-field components tend to exhibit larger Doppler velocities, with opposite signs on either side. This indicates rotational (vortex) motions along the thread-like magnetic field structure in the chromosphere.
%The relationship between the LOS magnetic field and the intensity or Doppler velocity is nearly identical in both \ion{Ca}{2} lines.

\subsection{ROI2: Filamentary Magnetic Field Structure Overlying the Internetwork}\label{sec:roi2}

\begin{figure*}[ht!]
\centering{\includegraphics[width=0.75\linewidth]{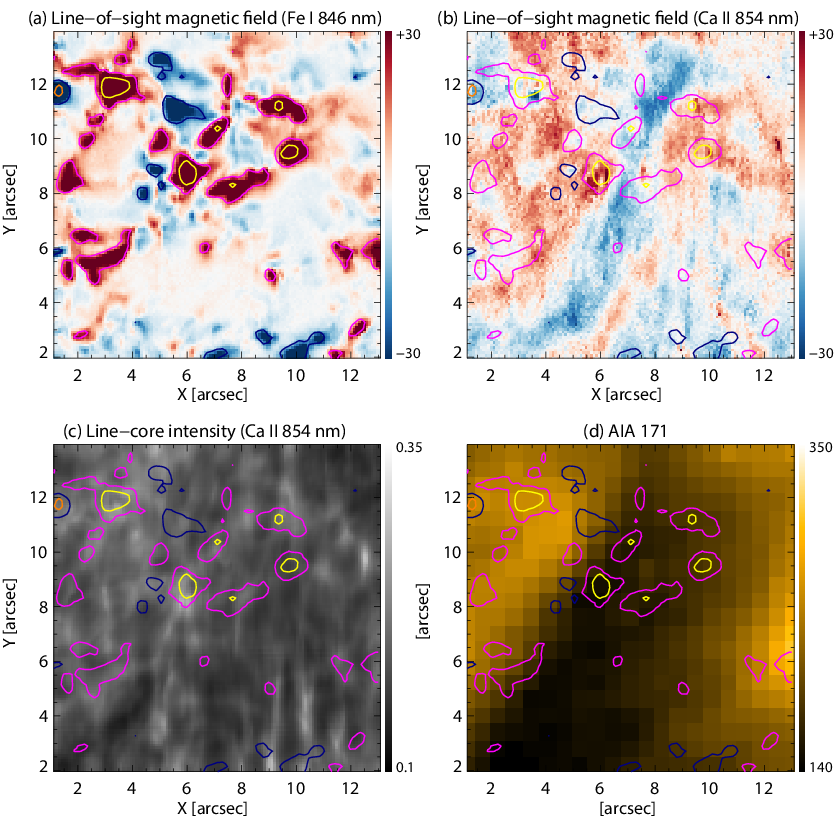}}
%\plotone{figures/apjl_SCIP_qs_ROI2_v21.eps}
\caption{
ROI2: 
(\textit{a})--(\textit{b}) LOS magnetic field for \ion{Fe}{1} 846~nm and \ion{Ca}{2} 854~nm.  
(\textit{c}) \ion{Ca}{2} 854~nm line-core intensity.  
(\textit{d}) SDO/AIA 171~{\AA} image (DN) at 06:09:45~UT. 
\lr{\lm{The} magenta, navy, yellow, and orange contours indicate values of $+20$, $-20$, $+100$, and $-100$~G, respectively.}
%Green arrows point at a dark blob.
\label{fig:ROI2}}
\end{figure*}

A second region (ROI2) reveals a filamentary chromospheric magnetic field structure of negative polarity situated between diffuse, positive-polarity canopy fields (Figure~\ref{fig:ROI2}(b)).
The photosphere beneath this structure is dominated by positive polarity overall, matching the diffuse chromospheric background. 
Mixed polarities are apparent beneath the upper portion of the filament, whereas no strong magnetic elements are present below its lower portion. 
This configuration differs from that of a typical dark filament, which generally forms above photospheric polarity inversion lines located between opposite-polarity magnetic elements.
\lr{The filamentary opposite-polarity structure seen in the chromospheric LOS magnetic field map (Figure~\ref{fig:ROI2}(b)) does not have a clear counterpart in the \ion{Ca}{2} 854~nm line-core intensity map (Figure~\ref{fig:ROI2}(c)).}
%A short\, dark blob is visible in the \ion{Ca}{2} 854~nm line-core intensity map, as indicated by the green arrow in Figure~\ref{fig:ROI2}(c).
%This blob is slightly offset from the opposite-polarity filamentary structure and is present only along part of it.
%The remaining portions exhibit intensities similar to those of the surrounding network where vertical acoustic wave patterns dominate. 
In contrast, a dark coronal feature is observed in the 171~{\AA} channel of the Atmospheric Imaging Assembly \citep[AIA;][]{2012SoPh..275...17L} along the filamentary LOS magnetic field structure (Figure~\ref{fig:ROI2}(d)).
This dark feature corresponds to the western end of the elongated feature indicated by the green arrow in Figure~\ref{fig:SDO}(c).
\lr{The corresponding coronal dark feature} is continuously observed \lr{in AIA 171~{\AA} images} throughout the entire SCIP scanning period.
%The AIA 304\,\AA\ image also shows reduced intensity in this region, although the elongated structure is less distinct than in the 171\,\AA\ image. This result is consistent with the absence of a clearly defined dark filament in the chromospheric observations.
\lr{The \lm{combined} chromospheric--coronal picture suggests that the filamentary chromospheric thread is embedded between more diffuse canopy fields with opposite polarity. 
This configuration is suggestive of a local dip in the canopy field, while the co-spatial dark feature in AIA 171~{\AA} may indicate denser plasma along the same structure.}

\section{Summary and Discussion} \label{sec:sum}
High-precision, multi-wavelength, full-Stokes spectropolarimetry from \textsc{Sunrise~iii}/SCIP reveals that the canopy is not a smooth, monotonically expanding magnetic fan emanating from network footpoints in the quiet Sun. 
Instead, we find multiple substructures in the chromosphere observed as numerous thread-like magnetic field structures: (i) alternating stronger and weaker LOS components of the canopy fields near network boundaries and (ii) a filamentary opposite-polarity thread embedded between diffuse canopy fields in the internetwork.
The LOS field strengths of the threads typically differ from the surrounding canopy by only 10--20~G, and SCIP sensitivity enables a detailed investigation of these structures.
Regarding (i), we detect enhanced linear polarization signals in the chromosphere \lr{near} the network side of the weak LOS component, \lm{together with weaker magnetic elements in the photosphere, some of opposite polarity}.
\lr{This pattern is reminiscent of} the spine and intraspine structures in the photosphere of sunspot penumbrae, where horizontal magnetic field components are embedded within relatively vertical magnetic fields.
In particularly clear cases, the canopy intraspines have opposite polarity to the surrounding network.
A three-dimensional radiative-MHD simulation of chromospheric jets with twisted magnetic field lines has reproduced thread-like magnetic field structures with polarity opposite to that of canopy fields extending from network regions \citep{2017ApJ...848...38I, 2023MNRAS.523..974M}.
In simulations of stronger magnetic field regions (i.e., enhanced network regions) using the recently developed MURaM-ChE code \citep{2022A&A...664A..91P}, opposite-polarity elongated magnetic field structures are often formed, and these structures are associated with the twisted magnetic flux ropes \citep{2026A&A...708A...3O, Ondratschek2026InPrep}.
A qualitative comparison between the enhanced network simulations and our observation by \citet{Ondratschek2026InPrep} shows that many aspects are consistent with our observational findings, supporting the idea that the quiet-sun magnetic canopy possesses a complex three-dimensional magnetic structure. 

The results from this study \lr{suggest that localized dips in the canopy field lines can form in the quiet-sun chromosphere, as illustrated by ROI1 and ROI2 in our FOV.} 
The localized canopy dips \lr{tend to be associated with relatively} dark features in the chromosphere or corona, suggesting a top-heavy configuration that becomes prone to the interchange instability. 
A model based on the magnetic Rayleigh-Taylor instability has been proposed to explain dip structures formed along arch-filament systems in emerging flux regions \citep{2005Natur.434..478I}.
This model also suggests that local dip structures produce heating phenomena through magnetic reconnection. 
The relatively stable, non-transient local dips discovered in the quiet-sun canopy may be explained by a similar mechanism. 
\lr{Such threads embedded within the surrounding canopy} can foster anti-parallel field components at their edges, and this creates favorable conditions for small-scale reconnection and impulsive heating. 
This scenario is analogous to penumbral microjets in sunspots, where interleaving spine/intraspine fields have been implicated in jet generation \citep{2007Sci...318.1594K}. 
The network peripheries in the quiet Sun are well known as source regions of various dynamic phenomena, including network jets \citep{2014Sci...346A.315T} and spicules \citep{2019Sci...366..890S}. 
While the SCIP data set used in this study does not provide the cadence necessary to follow rapid temporal evolution, it is essential to investigate the relationship between such phenomena and the \lr{localized substructures} within the canopy using higher-cadence data sets.

%% Please use the acknowledgment and contribution environments. This will 
%% be anonomyized when the "anonymous" style option is used. 
\begin{acknowledgments}
\textsc{Sunrise~iii} is supported by funding from the Max-Planck-Förderstiftung (Max Planck Foundation), NASA under Grants \#80NSSC18K0934 and \#80NSSC24M0024 (``Heliophysics Low Cost Access to Space'' program), and the ISAS/JAXA Small Mission-of-Opportunity program and JSPS KAKENHI Grant Numbers JP18H05234 and JP23K25916. This research has received financial support from the European Union's Horizon 2020 research and innovation program under grant agreement No. 824135 (SOLARNET) and No. 101097844 (WINSUN) from the European Research Council (ERC). It has also been funded by the Deutsches Zentrum für Luft- und Raumfahrt e.V. (DLR, grant no. 50 OO 1608). \lv{The Spanish contributions have been funded by the Spanish MCIN/AEI/10.13039/501100011033 under projects RTI2018-096886-B-C5, PID2021-125325OB-C5, and PID2024-156066OB-C5, and from "Center of Excellence Severo Ochoa" awards to IAA-CSIC (SEV-2017-0709, CEX2021-001131-S), all co-funded by "ERDF A way of making Europe". }
The research activities and the flight operation of the SCIP team members, R. T. Ishikawa, M. Kubo, Y. Kawabata, and T. Oba, have been supported by JSPS KAKENHI Grant Numbers JP23KJ0299, JP24K07105, JP23K13152, and JP21K13972, respectively.
CQN acknowledges support from Grants PID2022-136563NB-I00/10.13039/501100011033, and PID2024-156538NB-I00 and PID2024-156066OB-C55 funded by MCIN/AEI/10.13039/501100011033.
\end{acknowledgments}


\appendix

\section{Estimation of Physical Parameters}

\subsection{Doppler velocity} \label{sec:dp}
The line-center position is determined by a second-order polynomial fit to the five wavelength points around the minimum intensity of each spectral line.
\lr{The reference wavelength for the Doppler velocity is determined as} the line center position averaged over the full FOV.
The positive and negative Doppler velocities correspond to the redshift and blueshift, respectively.

\subsection{Total Circular and Linear Polarization} \label{sec:tcp}
The total circular polarization (TCP) and total linear polarization (TLP) are defined as
\lm{
\begin{equation}
\mathrm{TCP} = \frac{\displaystyle\sum_{i} w_i\, \displaystyle\frac{V(\lambda_0+\Delta\lambda_i)}{I_c}}{\displaystyle\sum_{i} |w_i|},
\label{eq:tcp}
\end{equation}
\begin{equation}
\mathrm{TLP} = \frac{\displaystyle\sum_{i} \sqrt{\left(\displaystyle\frac{Q(\lambda_0+\Delta\lambda_i)}{I_c}\right)^2 + \left(\displaystyle\frac{U(\lambda_0+\Delta\lambda_i)}{I_c}\right)^2}}{N},
\label{eq:tlp}
\end{equation}
where $\lambda_0$ denotes the line-center wavelength of the \ion{Ca}{2} 854~nm line or the \ion{Fe}{1} 846~nm line at each pixel, $\Delta\lambda_i$ denotes the wavelength offset of the $i$-th sample from the line center, and $I_c$ is the continuum intensity computed by averaging the spectral region from \lv{850.6 to 850.7~nm} (Section~\ref{sec:overview}).   
The weight $w_i$ in Equation~(\ref{eq:tcp}) is $+1$ for $-\lambda_1 \le \Delta\lambda_i \le -\lambda_c$ (blue wing), $-1$ for $\lambda_c \le \Delta\lambda_i \le \lambda_1$ (red wing), and $0$ otherwise, so that the sums in Equation~(\ref{eq:tcp}) effectively run over the two wing bands on either side of the line core. 
We adopt the parameter set $[\lambda_1, \lambda_c] = [35.6~\mathrm{pm}, 3.95~\mathrm{pm}]$ for both spectral lines.
In Equation~(\ref{eq:tlp}), $N$ is the number of samples with $-\lambda_1 \le \Delta\lambda_i \le \lambda_1$. 
}

%\begin{equation}
%\mathrm{TCP} = \frac{\displaystyle \int_{\lambda_0-\lambda_1}^{\lambda_0-\lambda_c} \frac{V(\lambda)}{\lr{I_c}}\,\mathrm{d}\lambda - \int_{\lambda_0+\lambda_c}^{\lambda_0+\lambda_1} \frac{V(\lambda)}{\lr{I_c}}\,\mathrm{d}\lambda}{\displaystyle \int_{\lambda_0-\lambda_1}^{\lambda_0-\lambda_c} \mathrm{d}\lambda + \int_{\lambda_0+\lambda_c}^{\lambda_0+\lambda_1} \mathrm{d}\lambda},
%\end{equation}
%\begin{equation}
%\mathrm{TLP} = \frac{\displaystyle \int_{\lambda_0-\lambda_1}^{\lambda_0+\lambda_1} \sqrt{\left(\frac{Q(\lambda)}{\lr{I_c}}\right)^2 + \left(\frac{U(\lambda)}{\lr{I_c}}\right)^2}\,\mathrm{d}\lambda}{\displaystyle \int_{\lambda_0-\lambda_1}^{\lambda_0+\lambda_1} \mathrm{d}\lambda},
%\end{equation}
%where $\lambda_0$ denotes the line-center wavelength of the \ion{Ca}{2} 854~nm line or the \ion{Fe}{1} 846~nm line at each pixel, \lr{and $I_c$ is the  intensity computed by averaging the spectral region from .5 to 848.6~nm (Section~\ref{sec:overview}).}  
%Line-center positions are determined by a second-order polynomial fit to the five wavelength points around the minimum intensity.  
%We adopt the parameter set $[\lambda_1, \lambda_c] = [35.6~\mathrm{pm}, 3.95~\mathrm{pm}]$ for both spectral lines.

\subsection{Line-of-sight Magnetic Field} \label{sec:wfa}
The line-of-sight magnetic field ($B_{\mathrm{LOS}}$) is inferred by applying the weak-field approximation (WFA) to the Stokes~$I$ and $V$ profiles \lm{\citep{1973SoPh...31..299L, 2004ASSL..307.....L}}.  
Following the method described in \lm{\citet{2003A&A...407..741D} and \citet{2009ApJ...700.1391M}}, we perform a linear least-squares fit:

\lr{
\begin{equation}
B_\mathrm{LOS} = \frac{\displaystyle \sum_{i} \frac{\partial I(\lambda_0+\Delta\lambda_i)}{\partial\lambda}\, V(\lambda_0+\Delta\lambda_i)}{4.66\times10^{-13}\, \bar g \, \lambda_0^2 \, \displaystyle \sum_{i} \left( \frac{\partial I(\lambda_0+\Delta\lambda_i)}{\partial\lambda} \right)^2},
\label{eq:wfa}
\end{equation}
where $\bar{g}$ is the effective Land\'e factor, $\lambda_0$ is as defined in Appendix~\ref{sec:tcp},
and $\Delta\lambda_i$ denotes the wavelength offset of the $i$-th sample from the line center.
The adopted wavelength ranges follow \citet{2024ApJ...960...26K}:
$\pm23.7$~pm for the \ion{Ca}{2} 854~nm and 849~nm lines,  
$\pm39.5$~pm for the \ion{Fe}{1} 846~nm line,  
and $\pm36.0$~pm for the \ion{K}{1} 766~nm and 769~nm lines.}
The WFA is applicable when the Zeeman splitting is smaller than the width of the spectral line.  
Because the Land\'e factor of the \ion{Fe}{1} 846~nm line is relatively large ($g=2.5$), the center-of-gravity method \citep[COG,][]{1979A&A....74....1R} is applied in regions where the WFA-inferred LOS magnetic field exceeds 100~G.  
Such strong-field areas are confined mainly to the cores of network regions or strong magnetic patches.  
All data points used in Figure~\ref{fig:ROI1_plot} exhibit LOS magnetic fields below 100~G.
Therefore, the choice of when to switch from the WFA to the COG does not affect our conclusions.

\lr{
The uncertainty in $B_\mathrm{LOS}$ is estimated through a noise simulation. 
We consider photon noise as the dominant noise source, with a polarimetric precision of $\sigma = 0.04\%$ of the continuum intensity $I_c$ (Section~\ref{sec:obs}). 
Random Gaussian noise of amplitude $\sigma$ is added independently to both Stokes $I$ and Stokes $V$ profiles.
The unperturbed Stokes $I$ profile is taken as the mean profile averaged over regions with TCP $< 0.0001$ (i.e., magnetically quiet areas), 
and the unperturbed Stokes $V$ is set to zero. 
The $B_\mathrm{LOS}$ is then derived from the noise-added profiles following Equation~(\ref{eq:wfa}). 
This process is repeated 10,000 times, and the resulting distribution of $B_\mathrm{LOS}$ values is 
fitted with a Gaussian function. The standard deviation of the fitted Gaussian gives the $1\sigma$ uncertainty in $B_\mathrm{LOS}$, 
which is $\sigma(B_\mathrm{LOS}) \approx 4.4$~G for the \ion{Ca}{2} 854~nm line.
}

%% For this sample we use BibTeX plus aasjournalv7.bst to generate the

\bibliography{kubo_ref}{}
\bibliographystyle{aasjournalv7}



\end{document}